\documentstyle[11pt]{article}
\def\dalemb#1#2{{\vbox{\hrule height .#2pt
        \hbox{\vrule width.#2pt height#1pt \kern#1pt
                \vrule width.#2pt}
        \hrule height.#2pt}}}
\def\square{\mathord{\dalemb{6.8}{7}\hbox{\hskip1pt}}}

 \def\bd{\begin{document}} \def\ed{\end{document}}
\def\ds{\documentstyle} \let\fr=\frac \let\bl=\bigl \let\br=\bigr
\let\Br=\Bigr \let\Bl=\Bigl
\let\bm=\bibitem
\let\na=\nabla
\let\pa=\partial \let\ov=\overline
\newcommand{\be}{\begin{equation}}
\newcommand{\ee}{\end{equation}}
\def\ba{\begin{array}}
\def\ea{\end{array}}
\def\ft#1#2{{\textstyle{{\scriptstyle #1}\over {\scriptstyle #2}}}}
\def\fft#1#2{{#1 \over #2}}
\def\del{\partial}
\def\sst#1{{\scriptscriptstyle #1}}
\def\oneone{\rlap 1\mkern4mu{\rm l}}
\newcommand{\ho}[1]{$\, ^{#1}$}
\newcommand{\hoch}[1]{$\, ^{#1}$}
\newcommand{\bea}{\begin{eqnarray}}
\newcommand{\eea}{\end{eqnarray}}
\newcommand{\ra}{\rightarrow}
\newcommand{\lra}{\longrightarrow}
\newcommand{\Lra}{\Leftrightarrow}
\newcommand{\ap}{\alpha^\prime}
\newcommand{\bp}{\tilde \beta^\prime}
\newcommand{\tr}{{\rm tr} }
\newcommand{\Tr}{{\rm Tr} }
\newcommand{\NP}{Nucl. Phys. }
\newcommand{\tamphys}{\it Center for Theoretical Physics,
Texas A\&M University, College Station, Texas 77843}

\newcommand{\auth}{H.
L\"u\hoch{\dagger}, C.N. Pope\hoch{\dagger},
E. Sezgin\hoch{\ddagger} }

\begin{document}
\begin{flushright}
\hfill{CTP-TAMU-40/95}\\
\hfill{Imperial/TP/95--96/11}\\
\hfill{hep-th/9511203}\\
\end{flushright}

\vspace{20pt}

\begin{center}
{ \large {\bf DILATONIC $p$-BRANE SOLITONS}}

\vspace{30pt}

\auth

\vspace{15pt}

{\tamphys}
\vspace{10pt}

K.S. Stelle\hoch{\sst\star}

\vspace{15pt}

{\it The Blackett Laboratory, Imperial College, Prince Consort Road, London
SW7 2BZ, UK}

\vspace{40pt}

\underline{ABSTRACT}
\end{center}

     We find new 4-brane and 5-brane solitons in massive gauged $D=6$, $N=2$
and $D=7$, $N=1$ supergravities.  In each case, the solutions preserve half
of the original supersymmetry.  These solutions make use of the metric and
dilaton fields only.  We also present more general dilatonic $(D-2)$-branes
in $D$ dimensions.

{\vfill\leftline{}\vfill
\vskip	10pt
\footnoterule
{\footnotesize
	\hoch{\dagger}	Research supported in part by DOE
Grant DE-FG05-91-ER40633 \vskip	-12pt}  \vskip	10pt
{\footnotesize
       \hoch{\ddagger} Research supported in part by NSF Grant PHY-9106593
\vskip	-12pt} \vskip 10pt
{\footnotesize
        \hoch{\sst\star} Research supported in part by the Commission of the
European Communities under contract SCI*-CT92-0789} }

\pagebreak
\setcounter{page}{1}

     The recent advances in the understanding of duality in string theories
have led to a resurgence of interest in supersymmetric $p$-brane solitons in
the supergravities that arise as the low-energy effective actions of
fundamental theories of extended objects.   These super $p$-branes may also
exist as fundamental objects in their own right. Their importance is based
on the speculation that they seem to participate in a web of
interconnections in the non-perturbative regime \cite{demo}. Further
evidence for such connections was recently found in ref.\ \cite{wit1} where
it was proposed that $D=11$ supergravity emerges as the strong coupling
limit of the $D=10$ type IIA string. This result suggests that the $D=11$
supermembrane \cite{bst} on $S^1$ may be equivalent to the $D=10$ type IIA
string if all perturbative and non-perturbative effects are taken into
account.

     Large classes of super $p$-brane solutions have been studied in the
type IIA, type IIB and heterotic strings, both in 10 dimensions and in lower
dimensions (see for example ref.\ \cite{dkl}).  In general these solutions make
use of the antisymmetric tensors and dilaton.  In particular, there exists
an elementary $(n-2)$-brane and a solitonic $(D-n-2)$-brane associated with
an $n$-index antisymmetric tensor field strength in $D$ dimensions.  The
general solutions of this type were obtained in ref.\ \cite{lpss}.   A
particular case that arises in some supergravity theories is when the theory
has a cosmological term of the form $\ft12 k^2 e^{a\phi}$ in the Lagrangian.
This term can be viewed as a special case of the general discussion above,
where the field strength is a 0-form.  When such a term is present in the
Lagrangian, it gives rise to a $(D-2)$-brane.  A recently studied example is
the 8-brane \cite{pw} in the massive type IIA supergravity in 10 dimensions
\cite{r10}.

     In this paper, we shall be concerned with supergravity theories that admit
such purely dilatonic $p$-brane solutions. One class of such theories is that
of the gauged supergravity theories (see for example ref.\ \cite{ss}).
All of these have a cosmological term, with a coefficient depending
on the gauge coupling, corresponding to a potential for the dilaton. Another
class of such theories is that admitting an independent potential term
associated
with a ``topological'' mass parameter. Some theories admit both types of
dilaton potential. In these circumstances, new kinds of $(D-2)$-brane
solution can occur. We shall primarily be concerned for our examples with
gauged
$D=7$, $N=1$ and gauged $D=6$, $N=2$ supergravities. The former admits a
topological mass term for a 3-index potential; the latter admits a mass term
associated with a massive 2-index potential. First let us consider the
$D=7$ theory, whose field content is $(e_{\sst M}^{\sst A}, \psi_{\sst M}^i,
B_{\sst{MNP}}, A_{{\sst M}i}{}^j, \lambda_i, \phi)$, where $i$ is an $SU(2)$
doublet index.  The bosonic Lagrangian is given by \cite{tvan}
\bea
e^{-1} {\cal L} &=& R - \ft12 (\del\phi)^2 -\ft12 m^2\,
e^{-\ft8{\sqrt{10}}\phi}
+ \sqrt{2} gm\, e^{-\ft3{\sqrt{10}}\phi} + \ft12 g^2\,
e^{\ft2{\sqrt{10}}\phi}
\nonumber\\
&& -\ft1{48} e^{-\ft4{\sqrt{10}} \phi} F_{\sst{MNPQ}}F^{\sst{MNPQ}} -
\ft14 e^{\ft2{\sqrt{10}}\phi} F_{\sst{MN}i}{}^j F^{\sst{MN}}{}_j{}^i
\label{boslag}\nonumber\\
&&+\ft1{96} e^{-1}\epsilon^{\sst{MNPQRST}} F_{\sst{MNPQ}}
F_{\sst{RS}i}{}^j A_{{\sst T}j}{}^i + \ft1{228} e^{-1}m\,
\epsilon^{\sst{MNPQRST}} F_{\sst{MNPQ}} B_{\sst{RST}}\ ,
\eea
where $e$ is the determinant of the vielbein, $m$ is the topological mass
parameter and $g$ is the gauge coupling constant.  If $m$ and $g$ are set
to zero, the Lagrangian (\ref{boslag}) reduces to the standard one for
$N=1$ ungauged supergravity.   This would admit standard elementary membrane
and solitonic string solutions using the 4-form field strength, and
elementary particle and solitonic 3-brane solutions using the 2-form field
strengths. Such solutions with the same isotropic ansatz for the metric and
antisymmetric tensor no longer occur when the parameters $m$ or $g$ are
non-zero.  However, as we shall now show, there are new types of solution
that describe supersymmetric five branes. These solutions use only the
metric and the dilaton, but not the antisymmetric tensors.

    The equations of motion for the metric and dilaton, which follow from
(\ref{boslag}) with the antisymmetric tensors set to zero, are given by
\be
\square \phi = S_1\ , \qquad R_{\sst{MN}} =
\ft12 \del_{\sst{M}} \phi \del_{\sst{N}}\phi + S_2\, g_{\sst{MN}}
\ ,\label{eqofmo1}
\ee
where
\bea
S_1&=&-\fft{4m^2}{\sqrt{10}}\, e^{-\ft8{\sqrt{10}}\phi} +
\fft{3gm}{\sqrt{5}}\, e^{-\ft3{\sqrt{10}}\phi} - \fft{g^2}{\sqrt{10}}\,
e^{\ft2{\sqrt{10}}\phi}\ ,\nonumber\\
S_2 &=&\fft{m^2}{10}\, e^{-\ft8{\sqrt{10}}\phi} -\fft{\sqrt{2}gm}5 \,
e^{-\ft3{\sqrt{10}}\phi} -\fft{g^2}{10}\, e^{\ft2{\sqrt{10}}\phi}
\ .\label{eqofmo2}
\eea
Note that $S_1 = \fft{\del}{\del \phi} S_2$.  The metric ansatz is given by
\be
ds^2 = e^{2A} dx^\mu dx^\nu\eta_{\mu\nu} + e^{2B} dr^2 \ ,\label{metricform}
\ee
where $A$ and $B$, as well as the dilaton $\phi$, are functions of $r$ only.
Substituting the metric ansatz into (\ref{eqofmo1}), we obtain
\bea
\phi'' + 6 A'\phi' - B'\phi' &=& e^{2B} S_1 \ ,\nonumber\\
A'' + 6{A'}^2 - A'B' &=& 6A'' - 6A'B' + 6 A'^2 +\ft12 \phi'^2 = e^{2B}
S_2\ , \label{eqofmo3}
\eea
These equations can be integrated to give rise to two first-order equations,
namely
\be
\phi' = \fft{4m}{\sqrt{10}}\, e^{-\ft4{\sqrt{10}} \phi + B} -
\fft{g}{\sqrt5}\, e^{\ft1{\sqrt{10}}\phi + B}\ ,\qquad
A' = \fft{m}{10}\, e^{-\ft4{\sqrt{10}}\phi  + B} + \fft{g}{5\sqrt{2}}\,
e^{\ft1{\sqrt{10}}\phi + B}\ .\label{firstorder}
\ee
In fact, as we shall see later, these two first-order equations are implied
by requiring that half the supersymmetry be preserved.  It is now
straightforward to solve the equations.  Noting that $B$ can be chosen
arbitrarily, by reparametrising $r$, we can simplify the equations by
setting $B=4A$.  We then find that the solution is given by
\bea
B&=&4A\ ,\qquad e^{\ft{3}{\sqrt{10}}\phi} = r\ ,\nonumber\\
e^{-4A} &=& \fft{4m}{\sqrt{10}}\, r^{-\ft13} -
\fft{g}{\sqrt{5}}\, r^{\ft43}\ .\label{solution1}
\eea

     Having obtained the bosonic solution, we shall now verify that it
preserves half of the original supersymmetry of gauged $N=1$, $D=7$
supergravity. The supersymmetry transformation rules for the fermionic
fields $\lambda_i$ and $\psi_{\sst M i}$, in a bosonic background in which
the antisymmetric tensor fields vanish, are given by \cite{tvan}
\bea
\delta \lambda_i &=& \fft{1}{2\sqrt2} \del_{\sst M}\phi
\Gamma^{\sst M} \epsilon_i
- \fft{m}{\sqrt5}\, e^{-\ft4{\sqrt{10}} \phi} \epsilon_i + \fft{g}{2\sqrt{10}}
\, e^{\ft{1}{\sqrt{10}}\phi} \epsilon_i\ ,\nonumber\\
\delta \psi_{\sst M i} &=& D_{\sst{M}} \epsilon_i -\fft{m}{20}\,
e^{-\ft{4}{\sqrt{10}}\phi} \Gamma_{\sst M} \epsilon_i -
\fft{g}{10\sqrt{2}}\, e^{\ft1{\sqrt{10}} \phi} \Gamma_{\sst M}\epsilon_i\ .
\label{susytranf}
\eea
Thus it follows from (\ref{firstorder}) that supersymmetry is preserved
provided that
\be
\epsilon_i = e^{\ft12 A}\epsilon_i^0\ ,\qquad
\Gamma_r \epsilon_i^0 = \epsilon_i^0\ ,\label{d7susy}
\ee
where $\epsilon_i^0$ is a constant spinor.  The 5-brane solution that we
have obtained therefore preserves half the supersymmetry.  The solution
contains two parameters $m$ and $g$.  When $g=0$, the metric can be
expressed as
\be
ds^2 = \rho^{\ft18} dx^\mu dx^\nu\eta_{\mu\nu} + d\rho^2\ ,
\label{mmetric}
\ee
and when $m=0$, the metric can be written as
\be
ds^2 = \rho^2 dx^\mu dx^\nu \eta_{\mu\nu} + d \rho^2\ ,
\label{gmetric}
\ee
where $\rho \propto r^{\ft43}$ for the first case and $\rho \propto
r^{-\ft13}$ for the second case.

     Now we turn to gauged $D=6$, $N=2$ supergravity.  The field content
is $(e_{\sst M}^{\sst A}, \psi_{\sst M}^i, B_{\sst{MN}}, B_{\sst M},$
$A_{{\sst M}}{}^{ij}, \lambda_i, \phi)$, where $i$ is a $US\!p(4)$ vector
index. The bosonic Lagrangian is given by \cite{r6}
\bea
e^{-1}{\cal L} &=& R - \ft12 (\del \phi)^2 -\ft12 m^2\, e^{-\ft3{\sqrt2}
\phi} + 2 g m\, e^{-\ft1{\sqrt2} \phi} + \ft12 g^2\, e^{\ft1{\sqrt2}\phi}
\nonumber\\
&& -\ft1{12} e^{\sqrt2 \phi} G_{\sst{MNP}} G^{\sst{MNP}} -\ft14
e^{-\ft1{\sqrt2}\phi} (H_{\sst{MN}} H^{\sst{MN}} + F_{\sst{MN}}{}^{ij}
F_{\sst{MN}ij}) \label{boslagd6}\\
&&-\ft1{16} e^{-1} \epsilon^{\sst{MNPQRS}}(G_{\sst{MN}}G_{\sst{PQ}}
+ m B_{\sst{MN}} G_{\sst{PQ}} +\ft13 m^2 B_{\sst{MN}} B_{\sst{PQ}}
+ F_{\sst{MN}}{}^{ij} F_{\sst{PQ}ij}) B_{\sst{RS}}\ ,\nonumber
\eea
where $G_{\sst{MNP}} = 3 \del_{[\sst M} B_{\sst{NP}]}$,  $G_{\sst{MN}} = 2
\del_{[\sst M} B_{\sst N ]}$ and $H_{\sst{MN}} = G_{\sst{MN}} + m B_{\sst
{MN}}$.   The relevant terms in the supersymmetry transformation rules for
the fermionic fields are given by \cite{r6}
\bea
\delta \lambda_i &=& \ft{1}{2\sqrt2} \Gamma^{\sst M}\del_{\sst M} \phi
\epsilon + \ft{1}{4\sqrt2}\, (g e^{\ft1{2\sqrt2}\phi} - 3m
e^{-\ft3{2\sqrt2}\phi})
\Gamma_7 \epsilon_i\ ,\nonumber\\
\delta \psi_{\sst M i} &=& D_{\sst M} \epsilon_i -\ft{1}{8\sqrt2}\,
(g e^{\ft1{2\sqrt2}\phi} + m e^{-\ft3{2\sqrt2}\phi})\Gamma_{\sst M}\Gamma_7
\epsilon_i\ .\label{susyd6}
\eea
Proceeding as in the case of $D=7$, we obtain the $4$-brane solution, given
by
\bea
B&=&3A\ ,\qquad e^{\ft{1}{\sqrt2}\phi} = r\ ,\nonumber\\
e^{-3A} &=& \fft{3m}{2\sqrt2}\, r^{-\ft12} - \fft{g}{2\sqrt2}\, r^{\ft32}
\ .\label{sold6}
\eea
It is easy to verify that this solution preserves half of the original
$D=6$, $N=2$ supersymmetry, with $\epsilon_i$ satisfying
\be
\epsilon_i = e^{\ft12 A} \epsilon_i^0\ ,\qquad
\Gamma_r\Gamma_7 \epsilon^0_i = \epsilon^0_i\ .\label{d6susy}
\ee
When $m$ or $g$ is zero, the metrics of the 4-brane solutions can be
simplified, and are given by
\bea
ds^2 = \rho^{\ft29} dx^\mu dx^\nu\eta_{\mu\nu} + d\rho^2\ ,\qquad
g=0\ ,\nonumber\\
ds^2 = \rho^2 dx^\mu dx^\nu \eta_{\mu\nu} + d \rho^2\ ,\qquad m=0\ ,
\label{mgmetric2}
\eea
where $\rho\propto r^{\ft94}$ for the first case and $\rho\propto r^{-\ft14}$
for the second.

      Having obtained new 5-brane and 4-brane solutions in $D=7$, $N=1$ and
$D=6$, $N=2$ gauged supergravities respectively, it is interesting to
examine the relation of the two solutions.  Since $D=6$, $N=2$ gauged
supergravity cannot be obtained from Kaluza-Klein dimensional reduction of
$D=7$, $N=1$ supergravity, the 4-brane solutions in $D=6$ with two
parameters $m$ and $g$ are in general not related to the 5-brane solutions
by dimensional reduction.   However when $m=0$, the two metrics of the 5-brane
and 4-brane solutions can be related by dimensional reduction.  To see this,
we note that whenever one dimensionally reduces a term of the form
$e^{-a\phi} F_n^2$ in a Kaluza-Klein theory (our case corresponds to $n=0$,
{\it i.e.\ }a 0-form field strength), the value of $\Delta$, defined by $a^2
= \Delta - 2(n-1) (D-n-1) /(D-2)$, is always preserved \cite{lpss}. The
$\Delta$ value for the pure $g$ potential is $-2$ for both $D=7$ and $D=6$
whilst the $\Delta$ value for the pure $m$ potential is 4 for $D=7$ and 2
for $D=6$. Thus the above metrics for $m=0$ could be related by dimensional
reduction, while the metrics for $m \ne 0$ can not. Similarly, the 5-brane
metric with $\Delta=4$ in $D=7$ could be related to the 8-brane metric in
$D=10$
\cite{pw} by dimensional reduction. The details of dimensional reduction
between supergravity theories with dilaton potentials of the types considered
in
this paper have not yet been worked out, however.

     Now we shall generalise these results and discuss the $(D-2)$-branes
in $D$ dimensional supergravity. In general, there are two different types
of $(D-2)$-brane solutions.  One of them is obtained by making use of a
cosmological constant term {\it i.e.\ }as we discussed previously.  In some
cases, the cosmological constant term can be described in terms of a
$D$-form field strength.   Under these circumstances, one can make use of
this field strength to construct an elementary $(D-2)$-brane solution.  In
fact the former description with the cosmological term can be viewed as a
solitonic solution using a 0-form field strength.  (However, note that in
the formulation using a cosmological term, the $(D-2)$-brane is necessarily
present, since Minkowski spacetime is not a solution of the equations of
motion.)  The metrics of these two types of $(D-2)$ branes are identical,
and hence without loss of generality, we can focus only on the
$(D-2)$-branes of the first type. Unlike the case of gauged $D=7$, $N=1$ and
$D=6$, $N=2$ supergravity, in general the supergravity theories constructed
so far admit no more than one cosmological term. The bosonic Lagrangian of
$D$ dimensional supergravity involving the metric, the dilaton and a
cosmological term is given by
\be
e^{-1}{\cal L} = R -\ft12 (\del\phi)^2 -\ft12 m^2 e^{-a \phi}\ ,
\ee
where $a$ can be parametrised as $a^2 = \Delta + \fft{2(D-1)}{D-2}$, and the
value of $\Delta$ depends on the specific supergravity theory.   It is a
simple matter to solve the field equations with the isotropic metric ansatz
(\ref{metricform}). In fact the solution is a special case of the
general results obtained for an $n$-form field strength
in ref.\ \cite{lpss}, in which we take $n=0$. The metric of the
$(D-2)$-brane solution is given by \cite{lpss}
\be
ds^2 = r^{\ft4{\Delta (D-2)}}\, dx^\mu dx^\nu \eta_{\mu\nu} +
r^{\ft{4(D-1)}{\Delta (D-2)}} dr^2\ .
\ee
which can be re-expressed as
\be
ds^2 = \rho^{\ft{4}{2(D-1) + \Delta (D-2)}} dx^\mu dx^\nu \eta_{\mu\nu}
+ d\rho^2\ ,\label{genmetric1}
\ee
where $\rho \propto r^{\ft{2(D-1)}{\Delta (D-2)} +1}$.  We expect that such
a solution preserves half of the supersymmetry.  An exception arises when
the dilaton is absent, which corresponds to $\Delta = -2 (D-1)/(D-2)$.   In
such a case the metric of the solution becomes $ds^2 = e^\rho\, dx^\mu
dx^\nu \eta_{\mu\nu} + d\rho^2 $, which is anti-de Sitter space, and hence
all of the supersymmetry is preserved.  It is easy to see that the metric
(\ref{genmetric1}) gives rise precisely to the metrics (\ref{mmetric}),
(\ref{gmetric}) and (\ref{mgmetric2}) when $D=7$ with $\Delta=4$ and $\Delta
=-2$ and $D=6$ with $\Delta =2$ and $\Delta=-2$ respectively. Other
examples include the ($\Delta=4$) 8-brane solution of the massive type IIA
supergravity in 10 dimensions \cite{pw}.  An example in which all the
supersymmetry is preserved is provided by the ``supermembrane'' in $D=4$,
$N=1$ supergravity with a cosmological term, where there is no dilaton field.

     Although only two supergravity theories constructed so far contain both
the mass and gauge parameters $m$ and $g$, it is nevertheless of interest to
obtain solutions in a generic dimension for a bosonic theory with the two
parameters.  If further supergravity theories with the two parameters exist,
then the results obtained below would be applicable. The relevant part of
the Lagrangian for such a theory is
\be
e^{-1}{\cal L} = R - \ft12 (\del \phi)^2 -\ft12 m^2\, e^{- a_1\phi} +
\ft12 g^2\, e^{a_2 \phi} + \lambda gm\, e^{\ft{a_2-a_1}{2} \phi}\ .
\label{boslaggen}
\ee
The equations of motion with the metric ansatz (\ref{metricform}) are given
by
\bea
\phi'' + (d A' -B') \phi' &=& S_1\ ,\nonumber\\
A'' + (dA' -B') A' &=& dA'' + d(A'-B')A' + \ft12 \phi'^2 = S_2\ ,
\label{eqmogen}
\eea
where
\bea
S_1&=& -\ft12 a_1 m^2\, e^{-a_1 \phi+2B} - \ft12 a_2 g^2\, e^{a_2 \phi+2B} -
\ft12 (a_2-a_1) \lambda g m \, e^{\ft12 (a_2-a_1) \phi+2B}\ ,\nonumber\\
S_2 &=& \fft{1}{2(D-2)} ( m^2\, e^{-a_1\phi} -g^2\,  e^{a_2\phi} -\lambda
g m \, e^{\ft12 (a_2-a_1)\phi} ) e^{2B}\ .\label{s1s2gen}
\eea
In order to solve the equations, we make a supersymmetry-inspired ansatz,
namely
\bea
\phi' &=& x_1 e^{-\ft12 a_1\phi+B} + x_2 e^{\ft12 a_2 \phi+B}\ ,\nonumber\\
A' &=& y_1 e^{-\ft12 a_1 \phi+B} + y_2 e^{\ft12 a_2 \phi+B} \ ,
\label{genansatz}\\
S_1 &=& \alpha \phi'^2 + \beta \phi' A'\ .\nonumber
\eea
Substituting this ansatz into the equations of motion (\ref{eqmogen}), we find
that we can solve for $x_1, x_2, y_1$, $y_2, \alpha$ and $\beta$ in terms of
the parameters $a_1, a_2, g, m$ and $\lambda$.  The expressions are very
complicated, and we shall not present them here.  However we note that there
are two distinct solutions.   The constants $x_1, x_2, y_1$ and $y_2$ take
different values in the two solutions, but $\beta$ and $\alpha$ are the same
for each solution.   In particular, the value of $\alpha$ is simple, given
by $\alpha = \ft12 (a_2 - a_1)$.  Choosing $B=(d-\beta) A$, we find that the
solutions for the $(D-2)$-brane take the form
\bea
B&=&(d-\beta) A\ , \qquad e^{\ft12 (a_1 - a_2)\phi} = r\ ,\nonumber\\
e^{-(d-\beta)A} &=& x_1\, r^{\ft{a_2}{a_2-a_1}} + x_2\,
r^{-\ft{a_1}{a_2-a_1}}\ .
\eea
The two solutions differ only in the values of $x_1$ and $x_2$.  In the
cases $D=7$ and $D=6$, one of the above solutions includes the
supersymmetric 4-brane and 5-brane that we obtained previously.  The other
corresponds to non-supersymmetric 4-brane and 5-brane solutions.

     In concluding, we would like to make some comments on the asymptotic
structure of the solutions (\ref{solution1},\ref{sold6}). $(D-2)$-brane
solutions in $D$ dimensions have only one transverse dimension, and so the
usual discussion of whether a BPS bound written in terms of an integrated
charge is saturated becomes somewhat awkward. Let us first consider the
question
of asymptotic flatness in the transverse direction. For the $D=7$ solution
(\ref{solution1}), obtained from our metric ansatz (\ref{metricform}) with
$B=4A$, we may write the curvature 2-form $\Theta^{MN}$ in terms of
basis 1-forms $(e^\mu=e^Adx^\mu, e^z=e^{4A}dr)$, finding
\bea
\Theta^{z\mu}&=&(-A''+3(A')^2)e^{-8A}e^z\wedge e^\mu\ ,\nonumber\\
\Theta^{\mu\nu}&=&-(A')^2e^{-8A}e^\mu\wedge e^\nu\ .\label{curv7}
\eea
For the $D=6$ solution (\ref{sold6}), with $B=3A$, in terms of the
basis one-forms $(e^\mu=e^Adx^\mu, e^z=e^{3A}dr)$ one finds
\bea
\Theta^{z\mu}&=&(-A''+2(A')^2)e^{-6A}e^z\wedge e^\mu\ ,\nonumber\\
\Theta^{\mu\nu}&=&-(A')^2e^{-6A}e^\mu\wedge e^\nu\ .\label{curv6}
\eea

     In order to find the flat regions, we need to discuss the cases $g=0$ and
$g\ne0$ separately. When $g=0$, the curvature for solution (\ref{solution1})
falls off as $r\rightarrow\infty$ like $r^{-8/3}$ in the 1-form basis and like
$r^{-3}$ for (\ref{sold6}). Thus, the $r\rightarrow\infty$ region is
asymptotically flat. In all cases discussed in this paper, the dilaton is
proportional to $\log r$. Whether this means that the flat space at transverse
infinity is ``empty'' or not is open to debate. The failure of the dilaton to
fall off is a direct consequence of the low codimension of the solution.

     Now consider the case $g\ne0$. In this case, the expressions
(\ref{curv7},\ref{curv6}) give a rising curvature as $r\rightarrow\infty$,
growing (in the chosen 1-form bases) like $r^{2/3}$ for (\ref{solution1}) and
like $r$ for (\ref{sold6}). Thus, the $r\rightarrow\infty$ region is not
asymptotically flat for this case. However, the region where $r\rightarrow 0$
is asymptotically flat in this case provided $m=0$. In both the $g=0$ and
$m=0$ cases, the proper distance along transverse geodesics between $r=0$
and $r=\infty$ diverges. Thus it emerges that provided {\it either}
$g=0$ {\it or} $m=0$, there is a flat region that may be considered to be
transverse infinity. In the $m=0$ case, a further coordinate transformation
$r\rightarrow 1/r$ relabels transverse infinity in the standard fashion. Note
that this flip does not destroy the proportionality between the dilaton and
$\log r$.

    We shall not proceed further here with a standard analysis of the BPS
bound. However, the saturation of this bound should be guaranteed by the
preservation of part of the original supersymmetry, as we showed in Eqs
(\ref{d7susy},\ref{d6susy}). The question of finding ``normalizable''
zero-modes describing the fluctuations about our BPS solutions is again
somewhat awkward here, without a convenient definition of the charges that
would normally be required to be convergent transverse integrals in higher
codimension cases. Nonetheless, one may identify appropriate worldvolume
supermultiplets that the zero modes could belong to. For the $D=7$ solution
(\ref{solution1}), the worldvolume is $d=6$ dimensional, and there is a natural
$d=6$ supermultiplet \cite{hsit} $(A^+_{\mu\nu},\lambda_{\alpha\,i},\varphi)$,
where the 2-form potential $A^+_{\mu\nu}$ has a self-dual 3-form field strength
$F^+_{\mu\nu\rho}$. The dimensional reduction of this multiplet to $d=5$ is
correspondingly available for the zero modes of the $D=6$ solution
(\ref{sold6}). Consequently, we may tentatively identify these multiplets as
the zero-mode supermultiplets.

\section*{Acknowledgement}

H.L., C.N.P. and K.S.S. thank SISSA, Trieste, and E.S. thanks ICTP, Trieste,
for hospitality during the course of this work. K.S.S. would like to thank
Arkadi Tseytlin for discussions.

\end{document}